\documentclass[pra,aps,showpacs,amsmath,amssymb]{revtex4}
\usepackage{graphics,epsfig,graphicx}
\usepackage{color}

\usepackage{subfigure}

\newcommand{\eq}{\begin{eqnarray}}
\newcommand{\en}{\end{eqnarray}}

\newcommand{\ud}[1]{{#1^{\dagger}}}

\newcommand{\be}{\begin{equation}}
\newcommand{\ee}{\end{equation}}
\newcommand{\bea}{\begin{eqnarray}}
\newcommand{\eea}{\end{eqnarray}}

\newcommand{\bra}[1]{\ensuremath{\langle #1 |}}
\newcommand{\ket}[1]{\ensuremath{| #1 \rangle}}

\begin{document}

\title{Crossover from Bosonic to Fermionic features in  Composite Boson Systems}
\author{A. Thilagam}

\affiliation{Information Technology, Engineering and Environment, 
Mawson Institute,
University of South Australia, Australia
 5095.} 
\begin{abstract}
We study the quantum dynamics of  conversion of composite bosons 
into fermionic fragment species  with increasing densities of bound fermion pairs using
the open quantum system approach. The  Hilbert space of $N$-state-functions is decomposed
 into  a composite boson subspace and an orthogonal fragment subspace  of
quasi-free fermions that  enlarges as the composite boson 
constituents  deviate  from  ideal boson  commutation relations.
The tunneling dynamics of coupled composite boson states in confined systems is examined,
and the appearance of exceptional points  under experimentally testable 
conditions (densities, lattice temperatures) is highlighted.
The theory is extended to examine the energy transfer between macroscopically
 coherent systems such as  multichromophoric macromolecules (MCMMs)
 in photosynthetic light harvesting complexes.
\end{abstract}


\maketitle

\section{Introduction}
Composite quasi-particle systems such as excitons (coherent superpositions of electrons and holes) 
display phase-space filling  effects when the mean separation
between particles becomes comparable to the exciton Bohr radius. This arises
due to the Pauli blocking  \cite{imam,thilapauli} of scattered fermionic particles
which constitute  ``cobosons" \cite{comb07,comb08} or composite bosons.
The  difference between composite bosons and 
elementary bosons can be seen  for instance, in the invariance of
Pauli scattering during  exchanges  between specific fermionic
species that can be correlated
via several routes \cite{comb08}.  Such interactions are  non-existent in a collection of ideal bosons 
with negligible  overlap, spin  and internal
structure. Unlike the   conventional  commutation relations satisfied by 
ideal boson operators of bound fermion pairs,  
composite bosons systems obey  a series of commutation relations \cite{comb10,comb11a} 
that are reflective of  the underlying fermionic structure of the constituent particles.
The excitonic polariton \cite{deng,lauss} is  one such example of  a composite boson system
in which the influence of the  composite-particle effect on many-body physics can be studied.
When the dimensions of quantum dot
excitonic polaritons are decreased, anti-bunching features  appear.
The system undergoes a crossover from bosonic to fermionic  display in optical features, 
characterized by a shift from the   Rabi doublet   to a Mollow triplet in the optical  spectrum \cite{lauss}.
In other systems, the  
deviations from the ideal boson features may influence the
magneto-association of atoms into molecules
via Feshbach resonances \cite{exptfe1,exptfe2}.

Recently, several studies have focussed on  the links between the
composite particle nature of bosons and entanglement effects \cite{Law,woot,avancini}
using the principles of quantum information theory. It was shown that effects of 
the  Pauli exclusion principle diminish when entanglement features
dominates \cite{Law,woot,Sancho,Rama,Gavrilik}.
Measures such as purity $P$ \cite{woot,Tichy} have been proposed to quantify
the strength of entanglement between the constituent fermions. While
the exclusion principle imposes the requirement that an antisymmetric state
vector be assigned to an identical group of fermions, the rules   becomes less stringent 
when the fermions become entangled and 
lose their fermion identity. The exact association between entanglement
and the  exclusion principle is not well understood as the latter implicates some 
degree of nonlocal interactions  for fermionic particles of the same state to display
anti-bunching behavior. Beyond  a critical dimension,
the tendency of similar fermions to ``avoid" each other is lost, with 
the problem  very much dependent on the system parameters such as dimensions and thermodynamic
factors. Currently,  very few details exists on the nature of the quantum correlation 
(whether classical or non-classical) that underpins the action of the exclusion principle in composite
boson systems.

In this work, we examine the conversion of composite bosons 
 to  orthogonal fermionic fragment states 
due to dominance of  Pauli exclusion at increasing densities of the correlated
boson system. The specific case of excitons is considered, and 
the ionic conversion of ${\rm e-h}$ $\to X$ is examined as an equivalent of the 
well known excitonic Mott transition where the excitonic 
system  ionizes due to space filling effects \cite{mott,zimm} at high densities, resulting
in  a free electron-hole plasma $X$. The situation in which
the bound exciton state vanishes and merges into a  continuum
of scattered states can also be considered as the Mott transition.
In this case, the scattered states  exist as quasi-free fermions.
In a recent study \cite{semkat}, the Mott effect was noted to occur 
when the excitonic Bohr radius became
 nearly the same as  the screening
length, that is, the  bound electron-hole pair states  co-existed
with the high-density electron-hole liquid phase.
 While  the interplay of several many-body effects are responsible for the Mott transition, 
questions remain as to whether
 the conversion from  the excitonic state to fermionic plasma-like phase proceeds
in a continuous manner.

Different timescales can be seen to operate in the composite boson-fermion system: 
the highly entangled elementary bosons which operate  at 
 fast  times compared to the scattering times specific to fermionic species.
To this end, the  composite bosons  can be analyzed via  the
open quantum system approach based on 
the  system-plus-reservoir model. The total Hilbert space is  divided into  subsystems
according to  different timescales and/or states which exist in distinct  phases.
The  Fock-Hilbert space of  state vectors  associated with a system of identical
bosons (fermions) is spanned by only symmetric (antisymmetric) functions.
The completeness theorem \cite{gross} applies
to many-particle state vectors in distinct Fock-Hilbert subspaces of antisymmetric
and symmetric particle functions. However  fluctuation in the number of interacting
particles occur when energy  dissipates from one  subspaces to another.
In this work, we consider that the  Hilbert space of state-functions is decomposed
 into  a composite boson subspace and an orthogonal fragment subspace  of
quasi-free fermions. The latter subspace  enlarges as the composite boson 
constituents  deviate  from  ideal boson  commutation relations 
 due to enhanced fermionic features. 
The quantum master equation approach
of the Lindblad form \cite{lind,gor} may be used to investigate composite boson systems,
however evaluation of the  density matrix of such high dimensional systems  
will present insurmountable challenges. Stochastic formalisms
such as the quantum trajectory method \cite{carmi,molmer} involving non-Hermitian terms which
cause  quantum jumps, coupled with 
the Feshbach projection-operator partitioning technique \cite{fesh}
may provide a viable route to  study the dynamics of composite bosons.
In this study, we employ the Green's function formalism \cite{lipp,suna} to examine the dynamics
of composite boson systems.

 A coupled system of  boson condensate and fragment states with variable  fermionic
character has  implications in the field of 
quantum information and processing.
Composite bosons can  be studied in the
 context of quantum tunneling of macroscopically
 coherent systems such as the 
double-well Bose-Einstein condensate. The double-well  condensate is a well-known
lattice system capable of variable controls \cite{ng,greiner},  exhibiting
a range of quantum effects such as  self-trapping, Josephson
 oscillations \cite{oscil}, and entanglement \cite{bose,entr}.
In this work, we  report on oscillations that occur in  the coupled composite boson system,
with testable predictions based on the
 quantum dot excitonic systems. These  results are extended to analyze  
multichromophoric macromolecules (MCMMs) in photosynthetic light harvesting
complexes. MCMMs are systems of great interest \cite{adolp,flem,silbey,recom,thilhs1} 
due to the appearance of long-lived quantum coherences, even at physiological temperatures.

This  paper is organized as follows. In Section \ref{cobo} we provide 
a brief review of cobosons or composite boson states and their Schmidt decomposition properties, along with
description of the  Schmidt number and purity measures. 
The occurrence  of the fermionic fragment state which lies orthogonal to the composite boson
is highlighted.  In Section \ref{altcobo}, a description of other 
alternative measures of deviations from ideal boson characteristics is provided, and
numerical estimates of the bosonic deviation measure  in quantum dots of varying size and fermion pair number
is provided. In Section \ref{master}, we introduce the 
open quantum system model and master equation  for the exciton-fermion system.
The importance  of  Non-Markovian dynamics due to the fermionic background is briefly
described in this Section. In Section \ref{tense}, the tensor structure of 
the many particle Fock-Hilbert space is examined for the $N$ excitonic bosons,
and the tunneling dynamics of composite bosons states is numerically examined
using  the decay branching ratio in Section \ref{tunn}. A discussion of the 
appearance of exceptional points under experimentally testable 
conditions is included as well. In Section \ref{lhs}, we examine the application of results
obtained in Section \ref{tunn} to  pigment protein complexes in light-harvesting systems,
and present our conclusions  in Section \ref{con}.

\section {Schmidt decomposition of  Cobosons or composite boson states}\label{cobo}

Following earlier  formalisms of cobosons comprising two fermions \cite{Law,woot,Combescot2011},
the coboson creation operator of distinguishable fermions in the Schmidt decomposition involving 
a single index is written as
\be
B^\dagger=\sum_{j=1}^J \sqrt{\lambda_{j}}\; a_{j}^\dagger b_{j}^\dagger,
\ee
 where  $\lambda_j$ are the Schmidt coefficients. $ a^\dagger_{j}$ and 
$ b^\dagger_{j}$ are different fermion creation operators in the Schmidt mode $j$, and
$J$ denotes the number of Schmidt coefficients \cite{schm}.
 The distribution of $\lambda_j$ is linked to a measure of entanglement,
which is provided via the Schmidt number  \cite{Law}
\be
\label{schnum}
{\cal K} \equiv 1/\sum\limits_{j=0}^{\infty} {\lambda _j^2 }
\ee
The Schmidt number ${\cal K}$  is the quantum counterpart to  the classical Pearson correlation coefficient,
and is an important entanglement measure \cite{ter,san} where 
large ${\cal K}$ indicates  high correlations and entanglement.
${\cal K}$ is also linked to another quantity known as the  purity $P$ \cite{woot}
via $P$=$\frac{1}{\cal K}$, and varies between zero and one.
In the case of two particles, $P = \hbox{Tr}\;\rho^2$, 
where $\rho$ is the density matrix of the examined particle. 

For two   identical particles,  (fermions or bosons),
the symmetrization postulate constraints a boson (fermion) state associated with
the system to be totally symmetric (antisymmetric) under permutation of the particles.
As a consequence,  the Schmidt decomposition of the state 
involves  more than one term, and  a bipartite state
of  two indistinguishable particles is generally considered entangled.
This highlights the importance of the Schmidt number in determining entanglement 
in quantum states of identical particles \cite{ghir}.
$B$ and $B^\dagger$ obey the non-bosonic commutation relation, $[B,B^\dagger] = 1 + s \Lambda$,
where $s=+1 (-1)$ if the two interacting particles are bosons (fermions),
and $\Lambda= \sum\limits_{j=0}^{\infty} {\lambda _j } \left( {a_j^{\dag} a_j + b_j^{\dag} b_j } \right )$.
 The state of $N$ composite bosons appear as 
\be 
\label{istateN}
\ket{N}=\frac{1}{\sqrt{\chi_N}}
 \frac{\left(  B^\dagger \right)^N }{\sqrt{N!}} \ket{0},
\ee 
where deviations from the ideal boson characteristics are contained
in the normalization term $\chi_N$ obtained using $\bra{N}N\rangle$=1.
The effectiveness of the  operator $B$ as 
a bosonic annihilation operator can be seen via the action of 
operator $B$ on state $\ket{N}$ \cite{Law}
\be
\label{frage}
B\ket{N}  = \alpha _N \sqrt N \ket{N - 1}
+ \ket{\mathcal{F}_N},
\ee
where $\alpha _N$ is a ideality   parameter and 
$\ket{\mathcal{F}_N}$ is the fragment state resulting from the non-ideal nature of
$\ket{N}$. $ \alpha_N  = \sqrt {\frac{{\chi_N }}{{\chi_{N - 1} }}}$,
which  in the case of an ideal composite boson yields the normalization ratio  $\chi_{N \pm 1}/ \chi_{N} \to 1$.
This ratio is seen as a measure of the degree of  ideal bosonization for a state of $N$ cobosons,
and appears  in the pair number mean value  corresponding to the state $\ket{N}$
as
 \be 
\langle   \hat{N} \rangle=\frac{\bra{N} \hat{N} \ket{N}}{\langle N  | N \rangle}
{=}1{+}(N{-}1)\frac{\chi_{N{+}1}}{\chi_N}
\ee
 and  also in the
 commutator mean \cite{Law}, 
\be \bra N [B, B^\dagger] \ket N = 2 \frac{\chi_{N+1}}{\chi_N} -1 
\ee
 A neat inequality involving the upper and a lower bound to the normalization ratio was determined 
as $1- P\cdot N \le \frac{\chi_{N+1}}{\chi_N} \le 1-P$ \cite{woot}.

The fragment states, $\ket{\mathcal{F}_N}$ remain orthogonal to $ \ket{N - 1}$,
yielding the correction factor \cite{Law,Combescot2011}
\be
\label{frag} 
\bra{\mathcal{F}_N} {\mathcal{F}_N} \rangle
= 1 - \frac{\chi_{N+1}}{\chi_N}
 - N \left( \frac{\chi_{N}}{\chi_{N-1}} 
- \frac{\chi_{N+1}}{\chi_N} \right). 
\ee  
The ratio $\chi_{N+1}/\chi_N$  is strictly non-increasing as $N$ increases \cite{woot}, 
and for small bosonic deviations such that $\chi_{N+1}/\chi_N \approx 1-\delta$, the last term
in Eq.~(\ref{frag}) can be dropped, $\bra{\mathcal{F}_N}\mathcal{F}_N \rangle $= $\delta$
and the commutator mean, $\bra N [B, B^\dagger] \ket N$ = $1- 2\delta$.
In the limits, $ \alpha_N  \to 1$,  $\delta  \to 0$, 
$\bra{\mathcal{F}_N}\mathcal{F}_N \rangle\to 0$.
The formation of the fermionic fragment can be compared to 
the formation of an  electron-hole plasma  when the density of a collection of 
correlated electron-hole is increased, giving rise to 
an enhancement in fermionic features. With increasing closeness of interacting paired fermions,
the  electron-holes pairs become unbound as is the case when lattice temperature is 
also increased. 
The state of $N$ composite bosons (see Eqs.~(\ref{istateN}))
with a  well-defined atom number  evolves into a mixture of lower number states
and fragment state $\ket{\mathcal{F}_N}$, 
characterized by the fidelity decay, $\gamma$. The role of 
$\gamma$ is significant,  due to its influence in a   Zeno-like mechanism where
repeated measurements halts further decay of the composite boson
states.

\section {Alternative measures of deviations from ideal boson characteristics}\label{altcobo}
\begin{figure}[htp]
  \begin{center}
\subfigure{\label{bs}\includegraphics[width=7cm]{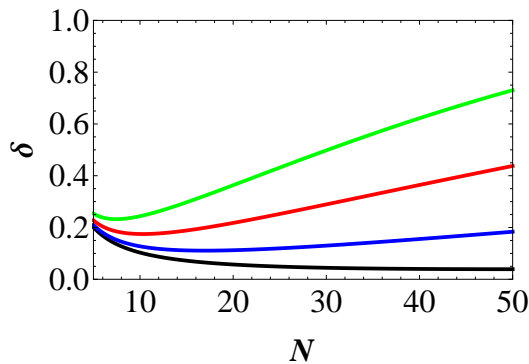}}\vspace{-1.1mm} \hspace{1.1mm}
     \end{center}
  \caption{Bosonic deviation measure $\delta$ as function of  fermion pair number $N$ in quantum dots at varying
ratios, $\frac{L}{a_\mathrm{B}}$= 0.01 (Dark solid line), 0.03 (Blue),
0.05 (Red),  0.07 (Green)}
\label{dev}
\end{figure}

Here we consider other measures that can be used in place of the purity $P$.
An experimentally accessible  measure that can be used to capture
 deviations from ideal boson characteristics is based on the 
(normalised) second order correlator~$g_2$ \cite{glaub,glaz}, 
which characterizes the probability of detecting of particles
at times $t$ and $t+\tau$ 
\be 
 \label{g2}
g_2(\tau)=\frac{\langle\ud{B}(t)\ud{B}(t+\tau)B(t+\tau)B(t)
\rangle}{\langle\hat N(t)\rangle\langle\hat N(t+\tau)\rangle}
\ee
$g_2(\tau)$ is based on 
correlations of the boson  operators where
$\hat N$ is the  number operator, $ \hat N\ket{n}=n\ket{N}$
associated with $N$ fermion pairs. At zero delay,
\be
g_2(0)= \frac{\langle N_0(t) (N_0(t)-1) \rangle}{\langle N \rangle^2}
\ee
where $\langle N_0(t)\rangle$ =$\langle\ud{B}(t)B(t)\rangle$.
The second-order correlator provides information 
on the underlying statistical features, such
as the Poissonian case ($g_2(0) = 1$) in coherent systems involving
a large number of Fock states. The  
 anti-bunching , sub-Poissonian case ($g_2(0) < 1$) is applicable in the fermionic limit at high
densities. The classically accessible thermal states
which display a bunched, super-Poissonian distribution 
gives rise to  $g_2(0) > 1$,  we do not consider such states in the work here.
A simple form of the zero-delay $\tau=0$ correlations for the single-mode
state, was obtained as $ g_2(0)=\frac{\alpha_{N-1}^2\alpha_N^2}{N^2}$ \cite{lauss},
where $\alpha_N = \sqrt{N}\sqrt{1- 2(N-1)(\frac{a_\mathrm{B}}{L})^2}$. The latter
expression is applicable to excitons of bohr radius $a_\mathrm{B}$ 
in  quantum dots of size $L$ at small values of $\frac{a_\mathrm{B}}{L}$
and  $n \ll \frac{L}{a_\mathrm{B}}$. In the pure bosonic case,
$g_2(0)=(N-1)/N \rightarrow 1$ as $N \rightarrow \infty$.
The  fermionic structure of excitons
becomes noticeable with decrease in $g_2(0)$, here we
estimate the bosonic deviations  using $\delta$ = $1- g_2(0)$.
Results displayed on Fig.~\ref{dev} show the increase in 
the bosonic deviation measure $\delta$ with 
increase in number density (or decrease in quantum dot size) as quantified by the 
ratio, $\frac{a_\mathrm{B}}{L}$. These results translates to the growth
of overlap in fragment states,
$\bra{\mathcal{F}_N}\mathcal{F}_N \rangle$=$\delta$
with increase in $\frac{a_\mathrm{B}}{L}$.

An alternative  measure of the bosonic deviations is obtained using 
the degree of binding of the paired fermions
\be \alpha_d= 1-\frac{E_c}{E_b} \ee
where $E_b$ is the maximized binding energy of the composite boson,
and $E_c$ is the smaller  binding energy of the quasi-bound fermion pairs,
which reaches zero for free fermions.
This definition is intimately linked to the distinguishability of 
paired fermions, a system of many strongly bound fermion pair is less distinguishable
and more entangled than one of free fermion species. In the case of 
excitons in  semiconductors at varying  temperatures and densities, a mixed phase
of bound excitons and free electron–hole plasma is formed. Here
the degree of  ionization  of the electron–hole plasma,  $ \alpha_i= 1-\frac{n_b}{n_f}$,
is a suitable candidate for gauging the bosonic deviations of the excitonic
system. The term $n_b$ denotes the excitonic density and $n_f$ denotes the density of the
free electrons or holes at a given temperature and density. The evaluation 
of $\alpha_i$ requires carrier density parameters such as 
the chemical potentials of distinct species, temperature, band
multiplicity,  spin degeneracy and an integrand with the
 retarded Green's function of carriers \cite{semkat}.

Yet another quantity that provides a measure of 
deviations from ideal boson characteristics is the $N$-particle non-escape probability in which 
$N$ paired fermions are found within the same region, $\Sigma$
\be
 P_N(t)=\int_{\Sigma} \prod_{n=1}^{N} dr_n |\Psi(r_1,...,r_N;t)|^2
 \ee
The evolution dynamics of  $P_N(t)$ can be  examined further by considering the tensor
structure of $P_N(t)$ in Fock-Hilbert space, however this procedure involves
the incorporation of all degrees of freedom of the $N$-particle system.
In Section \ref{tense}, we employ a similar approach, but one which requires  fewer parameters,
applied to the case of the excitonic bosons.

\section{Open quantum system model and master equation  for exciton-fermion system }\label{master}

We first examine the case of a paired electron-hole or exciton interacting with a background of 
dissociated electrons and holes. The total Hamiltonian of the exciton  and
fermion reservoir appear as
\bea
\label{THamilt}
\mathcal{H}_T &=& \mathcal{H}_{e}+
\mathcal{H}_{f}+\mathcal{H}_{i} +\mathcal{H}_{d},\\ \label{eHam}
\mathcal{H}_{e}&=&\sum_K(E_c- \mu_B)B_K^{\dagger}B_K \\ \label{ehHam}
\mathcal{H}_{f}&=&\sum _{k_{\rm e}}(\epsilon_{\rm e}-\mu_e) a_{k_{\rm e}}^\dagger \; a_{k_{\rm e}} 
+  \sum _{k_{\rm h}} (\epsilon_{\rm h}-\mu_h) h_{k_{\rm h}}^\dagger \; h_{k_{\rm h}} -
 \gamma_s \sum_{k,k'} a_{k}^\dagger h_{-k}^\dagger h_{-k'}a_{k'}
\\ \label{intHam}
\mathcal{H}_{i}&=&  \sum _{k_{\rm e},k_{\rm h}} \sigma^*(k_{\rm e},k_{\rm h}) a_{k_{\rm e}}^\dagger h_{-k_{\rm h}}^\dagger
B_{_{k_{\rm e}-k_{\rm h}}} + \sigma(k_{\rm e},k_{\rm h}) B_{_{k_{\rm e}-k_{\rm h}}}^\dagger a_{k_{\rm e}} h_{-k_{\rm h}}
\\ \label{dissHam}
\mathcal{H}_{d}&=& i \sum _{k_{\rm e},k_{\rm h}} \gamma_N^*(k_{\rm e},k_{\rm h}) B_{_{k_{\rm e}-k_{\rm h}}}^\dagger
B_{_{k_{\rm e}-k_{\rm h}}} 
\eea
where the subscripts, $e,f$ in the Hamiltonian terms refer, respectively to the exciton and  fermionic
fragments, $\mathcal{H}_{i}$ (Eq.~(\ref{intHam})) denotes the interaction
between the two subsystems.  $a_{k}^{\dagger}$ ($h_{k}^{\dagger}$) denote the 
electron (hole)  creation operator with wavevector $k_{\rm e}$($k_{\rm h}$)
and kinetic energy $\epsilon_{\rm e}$($\epsilon_{\rm h}$).
The boson creation operator, $B_K^{\dagger}$  is labeled by the wavevector $K$ 
and  is obtained using the Fourier transform of the site-dependent operator, $B_{ l}^\dagger$
as $B_{ K}^\dagger =  N^{-1/2} {\sum_{ l}} e^{i { K.l}} B_{ l}^\dagger$. 
The exciton creation operator $B_{ K}^\dagger$ which is localized in $K$-space,
is delocalized in real space.  $E_c$ in Eq.~(\ref{eHam}) may be seen as the minimum energy required to form the
composite boson system.
$\mu_e$ and  $\mu_h$ denote the respective chemical potentials for the electrons and holes,
while  $\mu_B$ denotes the chemical potential of the  bosonic system.
$\gamma_s$ denotes the strength of electron-hole interactions which occurs in the
fragment subspace. We  exclude, for simplification purposes, the background plasma of electron-hole
carriers that are originally formed when  excitons are created.
$ \sigma$ in the interaction operator $\mathcal{H}_{i}$ (Eq.~(\ref{intHam}))
is the momentum dependent coupling strength between two dissociated fermions and the correlated fermion
pair that constitutes the exciton. 
The non-Hermitian dissipative Hamiltonian $\mathcal{H}_{d}$ (Eq.~(\ref{dissHam}))
is dependent on  the  spontaneous decay  $\gamma_N$,  which increases with  greater deviations from ideal
bosonic features. This term may be attributed to
 the growth of the   free fermion plasma state, or fragment state (see Eq.~(\ref{frag})),
resulting  in the breakdown of boson states.

The density operator $\rho$ of the quantum system associated with the total 
 Hamiltonian, $\mathcal{H}_T$ (Eq.~(\ref{THamilt})) is obtained from  the generalized
Liouville-von Neumann equation  $\frac{d \rho}{dt} = -i {\cal L} \rho$. The generator
${\mathcal L}={\mathcal L}_b+{\mathcal L}_i+{\mathcal L}_f$  maps the initial to final density operators via a Liouville
superoperator $\Phi(t,0)$:   $\rho(0)\mapsto\rho(t)=\Phi(t,0)\rho(0)$. 
The  Liouville-von Neumann equation can be recast as a master equation of the 
following Lindblad form  \cite{lind,gor}:
\be
\frac{d}{dt}
\rho(t)= - i [\mathcal{H}_b+\mathcal{H}_i,\rho(t)] +  \sum_{k=1}^{d} {\gamma_k}
\left( {\cal V}_k\rho(t) {\cal V}_k^\dag-\frac{1}{2} \{{\cal V}_k^\dag {\cal V}_k,\,\rho(t)\}\right), 
\label{lind}
\ee
where  ${\cal V}_{k}$ denote Lindblad operators,
in which both an operator and its hermitian conjugate are labeled by $k$,
and the decay terms, $\{\gamma_k\}$ constitute the spectrum
of the positive definite $d$-dimensional Hermitian 
Gorini-Kossakowski-Sudarshan matrix ${\cal A}$   \cite{gor}.
The first term in Eq.~ (\ref{lind}) represents reversibility in system dynamics,
while the symmetrized  Lindblad operators, ${\cal V}_k$ involve transitions between
the many-particle levels of both the composite boson and  states present in the fermionic background.
While the notations associated with these possible energy transfer processes are not shown in 
 Eq.~(\ref{lind}), we note that these processes contribute to a 
 range of dynamical time scales due to  the possible
number of energetic degrees of freedom that can arise.  The multitude
of these transitions adds to the complexity of solving
many coupled differential equations of
Eq.~(\ref{lind}). The problem is however tractable in 
 systems with weak system-reservoir coupling when the  Markov approximation
applies. Next, we briefly describe the non-Markovian dynamics that may occur
in the composite boson system.

\subsection{Non-Markovian dynamics due to the fermionic background}\label{nonM}

At the initial period of quantum evolution,
the dynamics of the composite boson-fermion system is dominantly non-Markovian, determined
by the fermion  bath memory time, and the  Lindblad form in
Eq.~(\ref{lind}) does not provide a reliable description of the dynamics. Several processes may give rise to 
the non-Markovian dynamics, including the  dynamics associated with the vibrational 
environment of the composite boson-fermion system.
The time-scales of processes which occur in ideal and composite bosons
 differ  due to  decreased scattering between two elementary
bosons  as compared their composite counterparts \cite{combscat}.
 There also exist  differences between ideal bosons and composite bosons
in terms of the decreased scattering via lattice vibrations 
in non-ideal boson systems due to phase filling of the fermionic phase-space.

Information about the short time dynamics of the composite boson-fermion system
will help in the understanding of the subtle links between 
 non-Markovian dynamics  and  entanglement measures such as the Schmidt number ${\cal K}$  and purity $P$ 
of the parent composite boson state. This may  contribute to  
  accurate calculations of binding energies of complexes consisting of several fermions such as excitonic complexes \cite{thil1},
and improvements in density functional approaches to determining electronic properties of material systems. 
There are inherent difficulties in a detailed analysis of non-Markovian features, as this will require the  use of a non-Lindblad set of 
relations which incorporate  the finite time scale of the vibrational modes of the uncorrelated fermions.
To keep the problem tractable, we  consider in the next Section, the conversion of the composite boson state  $\ket{N}$
 to the fermionic fragment state  $\ket{\mathcal{F}_N}$ using
 a simple open quantum system consisting of
 a two-level  system  interacting linearly with a
dissociated electron-hole (fermion) reservoir.
We  note that as the density (or lattice temperature) increases,
the forward conversion of boson into free fermions is favored
 due to screening of  the
Coulomb interaction that tends to  bind and form composite bosons. Hence,
at higher densities, the interaction between the boson-fermion background
system is more likely to be  Markovian.

\section{Tensor structure of the many particle Fock-Hilbert space: $N$ excitonic bosons}\label{tense}
The tensor structure of
 the many particle Fock-Hilbert space for the system of $N$
correlated electron-hole or composite exciton state appear as
\be
\mathcal{S}_T = \mathcal{S}_{b}\otimes\mathcal{S}_{f}
\label{Hilbert}
\ee
where the total  many particle Fock-Hilbert space $\mathcal{S}_T$ is expressed as the tensor product of  two orthogonal subspaces. 
Here $\mathcal{S}_{b}$ ($\mathcal{S}_{f}$) is the subspace associated with the composite boson state  $\ket{N}$
(fragment  state, $\ket{\mathcal{F}_N}$ ). Both subspaces are considered to include the  complete set of bound and unbound states, and combination thereof to incorporate interactions present in any $N$-body system.

To keep the problem tractable, we assume that all $N$ bosons possess the same wavevector, $K=K_0$,
and consider  that the subspace 
$\mathcal{S}_{b}$ is  spanned by ($\ket{1}_b$, $\ket{0}_b$),
where $\ket{1}_b$  denotes the presence of $N$ composite bosons with wavevector, $K_0$,
and  $\ket{0}_b$ indicates a {\it lower} ($N' < N$) number of bosons
within $\mathcal{S}_{b}$. As noted in Section \ref{master}, the Fock space of composite boson and fermions
allows for many complicated interactions, coupled with the inseparability of the degrees of freedom
of the boson system and its fermionic background. The various correlations that occur
between $\ket{N'}$ and $\ket{N''}$ ($N',N'' < N$) is implicit in the defined $\ket{0}_b$,
and hence  we consider  a collective state that involves a superposition of similar states,
\be \ket{0}_b = \sum_{N' < N} A(N') \ket{N'} \ee
$A(N')$ is a weight factor that is dependent on the evolution dynamics of the coupled boson-fermion
system, and  for simplification we ignore details of the possible interactions between
the different composite boson states. The raising and lowering operators appear respectively, 
$\sigma^b_+$ =$ |1\rangle_b {}_b\langle 0|$ and
$\sigma^b_-$ =$|0\rangle_b {}_b\langle 1|$.
Due to the choice of definitions, a crude estimate of the 
energy difference between $\ket{1}_b$ and $\ket{0}_b$
states is given by the binding energy of a single composite boson.
This quantity  can be  obtained using experimental techniques in the case of excitonic systems.

Likewise, we also consider that   $\mathcal{S}_{f}$ is 
 spanned by ($\ket{1}_f, \ket{0}_f$) where 
$\ket{1}_f$  ($\ket{0}_f$) denotes the presence (absence) of the fragment state,
 $\ket{\mathcal{F}_N}$. Instead of identifying electron
and hole states, we denote  both types of fermions using operators $c_k^\dagger,c_k$,
and the raising  fermionic operator, $\sigma^f_+$ appear as 
linear combination of $c_k^\dagger$ operators
 \be 
\sigma^f_+ = |1\rangle_f {}_f\langle 0|= \frac{1}{\sqrt{A}} 
\sum_k j_k c_k^\dagger
\ee
which can be easily shown to obey the anti-commutator relation $\{\sigma^f_-,\sigma^f_+\}$ =1,
where the lowering operator, $\sigma^f_-$= $|0\rangle_f {}_f\langle 1|$.
$A$=$\sum_k \gamma_k^2$, where 
 $j_k$ is  the weight amplitude for the respective fermion in simplified
system of noninteracting fermions.

Using Feshbach  projection-operator partitioning method \cite{fesh},
the total Hilbert space of  $\mathcal{H}_T$ (Eq.~(\ref{THamilt}))
is  divided into two orthogonal subspaces, $\mathcal{S}_{b}$ and $\mathcal{S}_{f}$ (see Eq.~(\ref{Hilbert}))
generated respectively  by  a projection operator, ${\cal P}$=$ \sigma^b_+\sigma^b_-$ and its complementary,
 ${\cal Q}$=$\sigma^f_+\sigma^f_-$, with 
Hence ${\cal P Q}$=${\cal Q P}$=0, and the reduced density state operator associated with the central
composite boson system of interest is  obtained via
\bea
\label{proj}
\rho_b&=&{\cal P} \; \rho_T {\cal P}, \\ \label{trace}
&=&{\rm Tr_f} \{\rho_T(t)\}
\eea
where $\rho_T$ is the density operator of the total system described by
$\mathcal{H}_T$.
The reduced density state, $\rho_b$ can also be obtained by taking
the partial trace over the fermionic environment as shown in Eq.~(\ref{trace}).
The density operator associated with 
the fermionic fragment can be obtained using $\rho_f(t)={\cal Q} \; \rho_T {\cal Q}$.
If we consider that at time $t_0$, the subsystems of composite bosons
and fermionic fragment are in separable states,  the evolution of
$\rho_b(t)$ in Laplace space becomes
\be \rho_b(z)-[z-({\mathcal L}_b+N(z))]^{-1} \ee
where the energy term $N(z)$ is associated with non-Markovian interactions.
Feshbach  projection-operator partitioning has been employed in an earlier
work \cite{thilush}, to examine the
dynamics of open quantum systems  via the stochastic
quantum trajectory approach. While the stochastic route
presents a physical interpretation of the quantum trajectories in the case
of Markovian dynamics, it offers no viable explanation
for the occurrence  of non-Markovian dynamics,
due  to the finite correlation time
of the non-Markovian reservoir \cite{manisca}.
An alternative approach involving the 
 post-Markovian master equation \cite{lidar,manisca} is known
to be applicable  in the regime between
 Markovian and non-Markovian quantum 
dynamics in open quantum systems. Here we utilize the
Green's function approach \cite{lipp,suna} to  provide an  effective description  of the
quantum evolution of  the composite boson system.

\section{Tunneling dynamics of composite bosons states: appearance of exceptional points}\label{tunn}

We consider the tunneling dynamics between a pair of $N$-composite boson states  $\ket{1}_{b1}$
and $\ket{1}_{b2}$ with a  total Hamiltonian of the form ($\hbar$=1)
\be
\label{pheq}
\mathcal{H}_T = \omega_{b1} \; \sigma^{b1}_+\sigma^{b1}_- + \omega_{b2} \; \sigma^{b2}_+\sigma^{b2}_- 
+V^* \; \sigma^{b1}_+\sigma^{b2}_- + V\; \sigma^{b2}_+\sigma^{b1}_-
- i\gamma_{d1}  \sigma^{b1}_+\sigma^{b1}_-- i\gamma_{d2}  \sigma^{b2}_+\sigma^{b2}_-
\ee
where $\omega_{bi}$ ($i$=1,2) are the two  composite boson  
transition energies (c.f. Eq.~(\ref{eHam})). $V$ denotes  the 
tunneling energy between the two $N$-composite boson states, and is
taken to be real and positive, without loss in generality.
Due to the choice of definitions for $\ket{1}_b$ and $\ket{0}_b$,
the tunneling energy $V$ could  involve the transfer of excitation
associated with just one fermion pair. As will be described in 
Section \ref{lhs}, $V$ could represent a series of repeated processes
leading to transfer of states from one site to another in photosynthetic 
protein complexes. The dissipative terms in Eq.~(\ref{pheq})
represent  leakages of boson states into the fermionic subspaces,
 $\mathcal{S}_{b} \to \mathcal{S}_{f}$. 
The state $\ket{1}_{b1}$($\ket{1}_{b2}$)decays  at the rate $\gamma_{d1}$ ($\gamma_{d2}$) 
to the lower state $\ket{0}_{b1}$ ($\ket{0}_{b2}$).
The phenomenological rate $\gamma_{di}$=$\Delta_i \; \delta_i$, is employed 
where $\delta_i$=$\bra{\mathcal{F}_N}\mathcal{F}_N \rangle$
(see Eq.~(\ref{frag})) is a measure of 
the deviation from bosonic  features with increase in density of the composite
boson state in a specific subspace $\mathcal{S}_{bi}$.  $\Delta_i$ is taken as a constant  with units of energy.

We consider the retarded Green's function of the form
\be
G_{1,2}(t)=-i\Theta(t)\langle \{\sigma^{b1}_-(t),\sigma^{b2}_+(0)\}\rangle
\label{geq}
\ee
where $\Theta(t)$ denotes a step function. The Fourier transform, $G_{1,2}(E)$=$\int_{-\infty}^{\infty} dt\, G_{1,2}(t)\, e^{iEt}$
for the system in Eq.~(\ref{pheq}) appear with the Lippmann-Schwinger matrix terms \cite{lipp}
\be
\label{hermg}
G_{1,2}^{-1}(E)= \left[ \begin{array}{cc} E-\omega_{b1}+i \eta+ & -V \\ -V & 
E-\omega_{b2}+i \eta+ \end{array} \right]+ \frac{i}{2} \left( \begin{array}{cc} \gamma_{d1} & 0 \\ 0 & 
\gamma_{d2} \end{array} \right).
\ee
where $\eta$ is a very  small number and the dissipative process
associated with $\gamma_{d1}$ ($\gamma_{d2}$) in the subspace $\mathcal{S}_{b1}$ ($\mathcal{S}_{b2}$) is considered
as an  irreversible loss of the composite exciton of size $N$. 
In an initial state, $t=0$ and the  fermionic fragment will be absent,
and only the   $N$-state composite boson state  $\ket{1}_{b1}$
is excited. We denote the 
probability  of excitation to remain at its initial site, $P_{1,1}(0)$=1.
The probability $P_{1,2}$  of excitation transfer from one boson subspace to another, $\mathcal{S}_{b1} \to \mathcal{S}_{b2}$
is determined by inverting Eq.(\ref{hermg})
\be
P_{1,2}(t)=\frac{2V^2}{|\Omega|^2} e^{-\gamma_d t} 
(\cosh\Omega_i t - \cos\Omega_r t),
\ee
where $\gamma_d$=$\frac{1}{2}(\gamma_{d1}$+$\gamma_{d2})$, $\Omega \equiv \Omega_r + i\Omega_i\equiv\sqrt{4V^2 + 
(\omega_0-i \bar{\gamma_d})^2}$, $\omega_0$=$\omega_{b2}$-$\omega_{b1}$,$\bar{\gamma_d}$=
$\frac{1}{2}(\gamma_{d2}$-$\gamma_{d1})$.
In the absence of the Pauli exclusion related dissipation at time $t > t'$,
the boson-fermion system undergoes Rabi-type oscillation  determined by  $\omega_0$
and $V$. In the continued  presence of dissipative terms,
 the total probabilities, $P_{1,1}$+$P_{1,2} \le 1$ is not conserved and there is 
  loss of normalization which is  dependent on  $\gamma_{d1},\gamma_{d2}$.

\begin{figure}[htp]
  \begin{center}
\subfigure{\label{aa}\includegraphics[width=4cm]{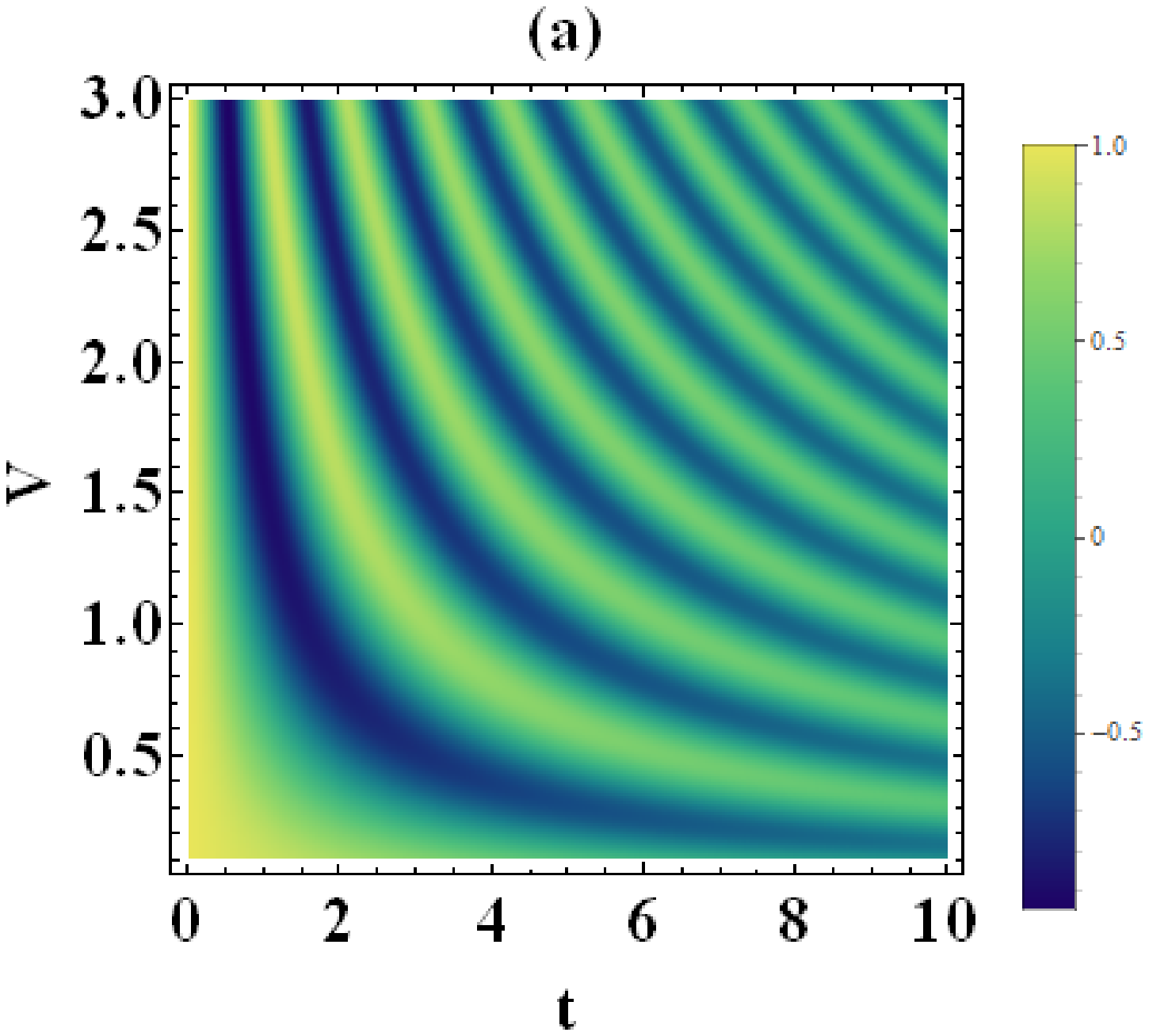}}\vspace{-1.1mm} \hspace{1.1mm}
\subfigure{\label{ab}\includegraphics[width=4cm]{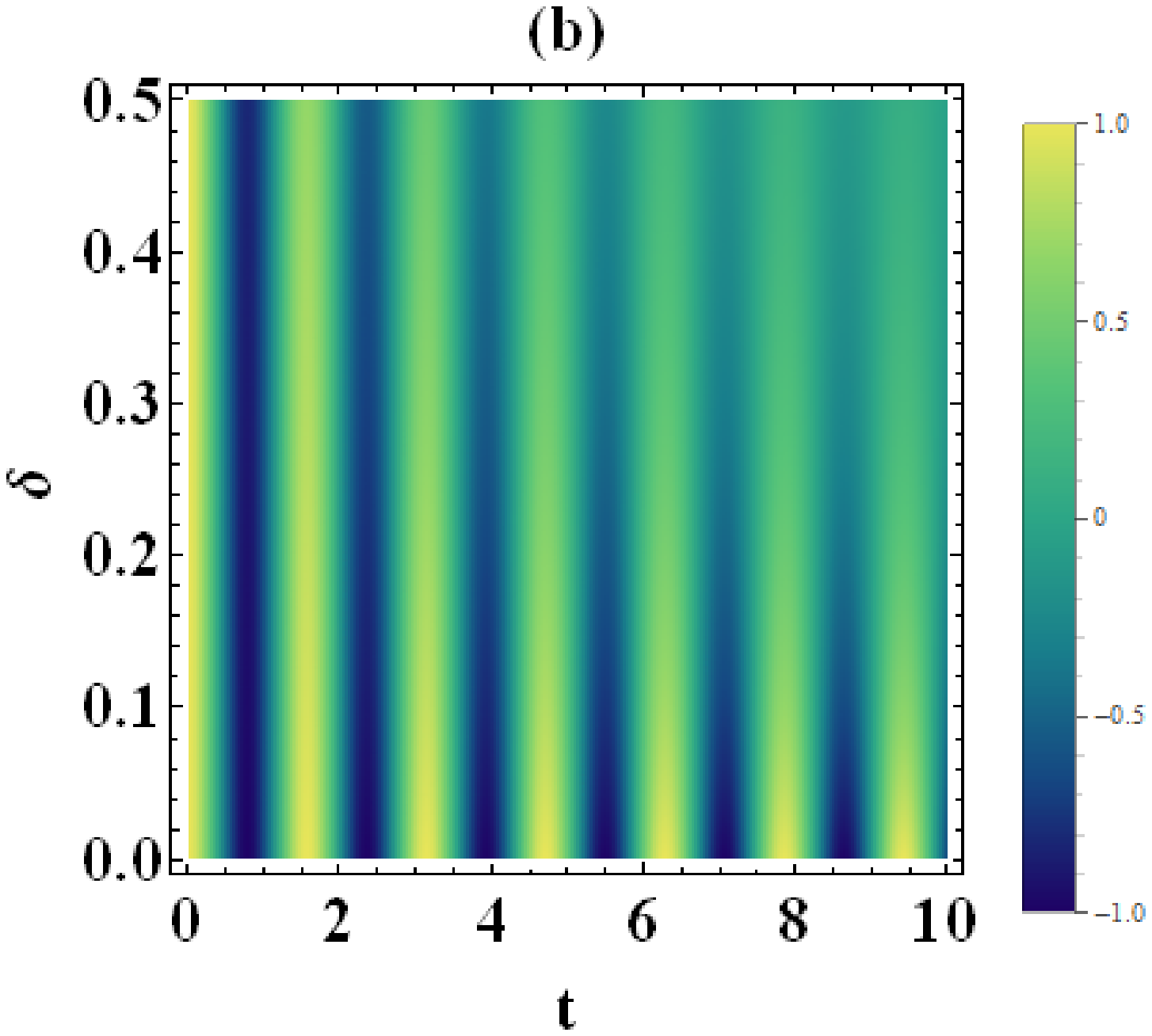}}\vspace{-1.1mm} \hspace{1.1mm}
     \end{center}
\caption{(a)Population difference, $\Delta P$=$P_{1,1}$-$P_{1,2}$ 
as a function of time $t$,  and  coupling energy, $V$ at dissipation rates 
$\gamma_{d1}=\gamma_{d2}$=0.1, for the degenerate case ($\omega_{b2}$=$\omega_{b1}$). 
The units are chosen such that $\hbar$=1, $\Delta_1$=$\Delta_2=1$ (i.e $\delta_1$=$\delta_2$=0.1).
 Time $t$  is obtained as multiple of $t_0$, the  inverse of 
 $\Omega_0$ (at $\gamma_{d1}=\gamma_{d2}$=0.1)  
(b)  Population difference, $\Delta P$=$P_{1,1}-P_{1,2}$ 
as a function of time $t$,  and  deviation factor, $\delta=\delta_2$ at 
tunneling energy $V$=1, $\gamma_{d1}$=0 and $\Delta_2$=1.
}
\label{tunnel}
\end{figure}

Fig.~\ref{tunnel}a,b shows the tunneling dynamics  for the specific case when the energy difference
$\omega_0$=0. 
Depending on the tunneling energy $V$ and  decay rates $\Gamma'$, there is 
existence of  a coherent regime ($2V > \bar{\gamma_d}$) or incoherent regime ($2V < \bar{\gamma_d}$).
At large $V$  and small dissipation levels, there is increased
exchanges between the two coupled bosonic states (Fig.~\ref{tunnel}a). 
The gradual decrease in the $N$-state  boson population appears
due to increased deviation from ideal boson characteristics
associated with increase in number density (or decrease in quantum dot size) as 
shown in Fig.~\ref{tunnel}b.

The appearance of  topological defects known as  exceptional points \cite{Heiss} occurs when
$ \Omega = 0, \; V = \frac{\bar{\gamma_d}}{2}$
Unlike degenerate points, only a single eigenfunction exists at the exceptional point due to the 
merging of two eigenvalues. The critical boson densities  and
lattice temperatures at which exceptional points
occur is evaluated using many body   theory which takes
 into account  dynamical screening processes. This screening may  arise from  Coulomb
interactions of the one-particle and two-particle properties between the same
and different fermion species constituting the composite boson system.
These special points may be
associated with a range of system parameter attributes, and hence lie within an allowed
spectrum that may be amenable to experimental detection.

\begin{figure}[htp]
\subfigure{\label{A}\includegraphics[width=4cm]{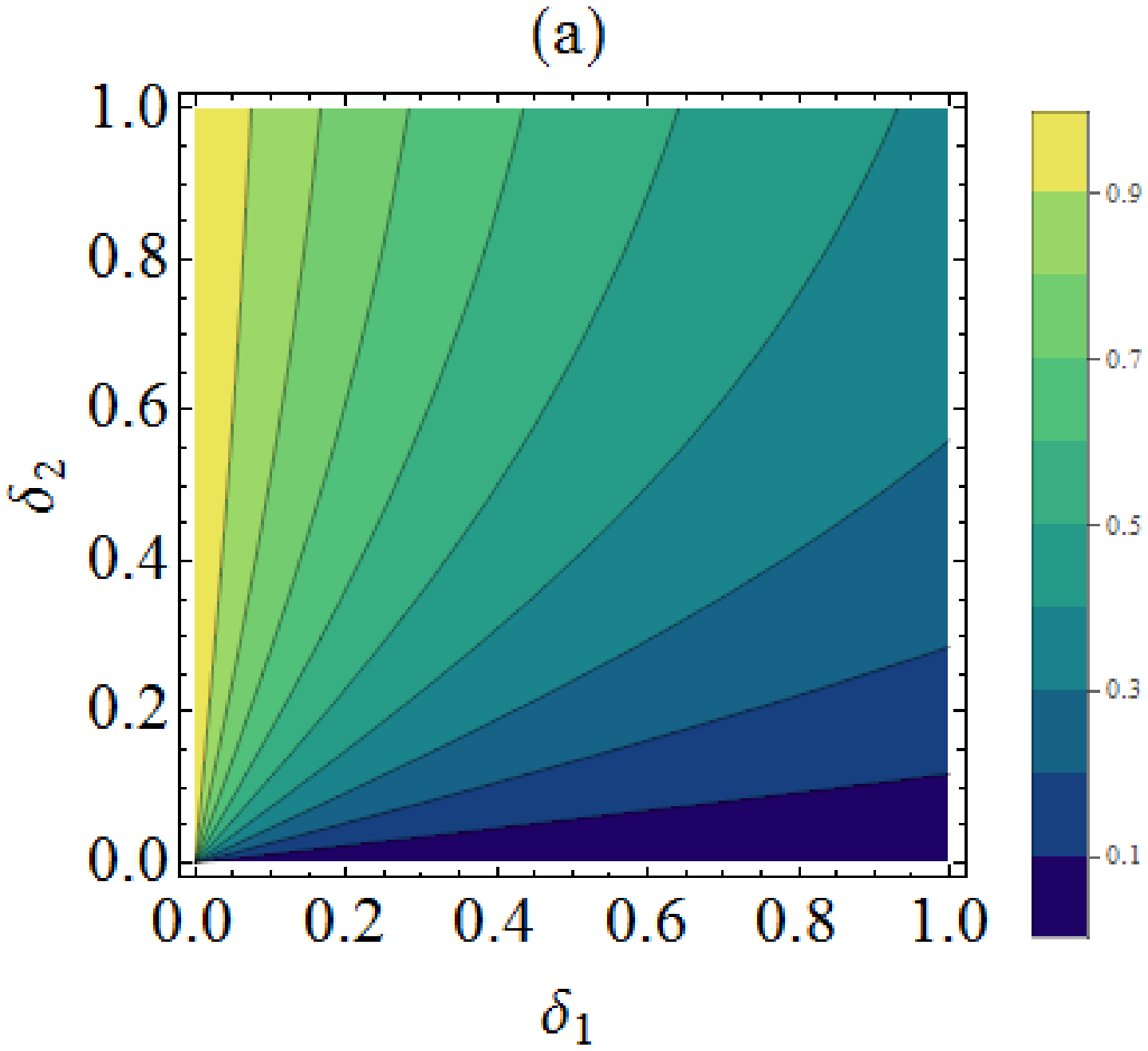}}\vspace{-1.1mm} \hspace{1.1mm}
\subfigure{\label{B}\includegraphics[width=4cm]{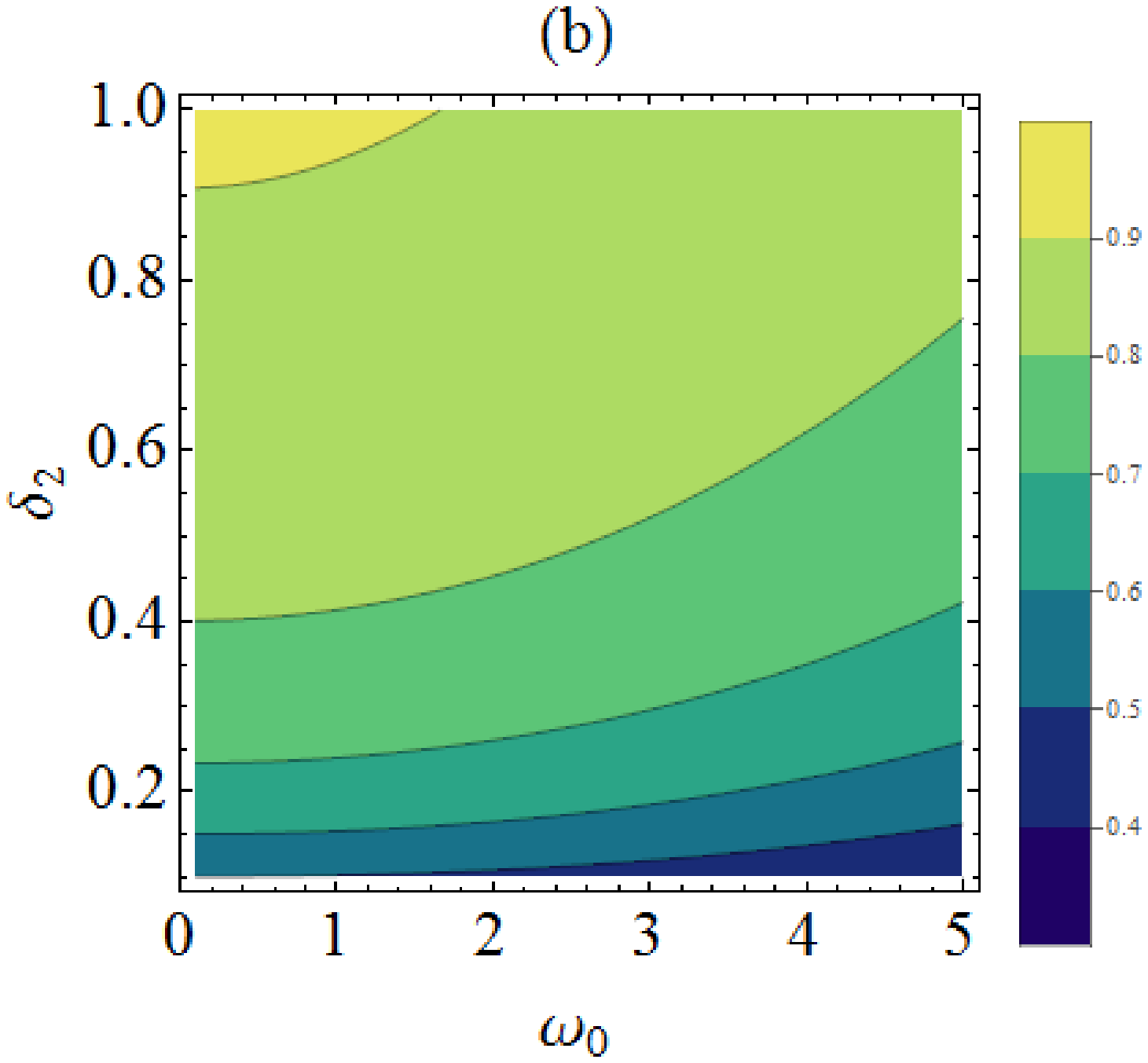}}\vspace{-1.1mm} \hspace{1.1mm}
\caption{(a) Branching fraction $F_2$ of a $N$-state composite boson that decay via $\gamma_{d2}$
as a function of bosonic deviations, $\delta_1$ and $\delta_2$. Energy difference $\omega_0$=$\omega_{b2}$-$\omega_{b1}$=0.5,
and tunneling energy $V$=1. 
(b) Fraction $F_2$ as a function of energy difference $\omega_0$ and bosonic deviation, $\delta_2$.
The tunneling energy $V$=5 and $\delta_1$=0.1.
The units are chosen such that $\hbar$=1, $\Delta_1$=$\Delta_2=1$ for both figures.}
\label{branch}
\end{figure}

The decay branching ratio  is quantified by the 
fraction $F_1$ (or $F_2$) of a $N$-state composite boson that decay via $\gamma_{d1}$ (or $\gamma_{d2}$). 
$F_2$ is evaluated using Parseval's theorem \cite{statt} 
\bea
F_2 &=&\gamma_{d2} \;\int_{-\infty}^{\infty} \frac{dE}{2\pi} |G_{1,2}(E)|^2,\\
&=&
 \frac{(1+\frac{\gamma_{d2}}{\gamma_{d1}})V^2}{{\omega_0}^2 + \gamma_{d}^2
(1+\frac{4V^2}{\gamma_{d2}\gamma_{d1}})}.
\label{pas}
\eea
The  fraction $F_1$=$1-F_2$. 
Results displayed on Fig.~\ref{branch}a show the increase (decrease) in 
the branching fraction $F_2$ due to increase (increase) in 
the bosonic deviation measure $\delta_2$ ($\delta_1$) for fixed values of
the energy difference $\omega_0$ and tunneling energy $V$.
 Fig.~\ref{branch}b shows the notable decrease of the branching fraction $F_2$
with increase in  the energy difference $\omega_0$. Conversely, 
the branching fraction $F_1$ increases with increase in  the energy difference $\omega_0$,
as expected.

\section{Application to pigment protein complexes in light-harvesting systems}\label{lhs}

It is useful to analyze the results obtained in Section \ref{tunn} in  the context
of   large photosynthetic membranes which constitute  many 
biological pigment-protein complexes (i.e., chromophores) such  as  FMO (Fenna-Matthews-Olson) complexes
in the green sulphur bacteria \cite{adolp,flem}. The FMO complex trimer is made up of  three symmetry equivalent
 monomer subunits, with each unit constituting eight bacteriochlorophyll
(BChl)a molecules supported by a cage  of protein molecules. 
The FMO  complex acts as an efficient channel  of  excitation transfer in which
 photons captured in the chlorosome which is  the main light harvesting  antenna complex,
 are  directed via a series 
of  excitonic exchanges to a reaction center (RC) where energy conversion  into a
 chemical form occurs.  Long-lived coherences between electronic states   lasting 
several picoseconds, much shorter than the $1$ ns dissipative  lifetimes  of excitons \cite{flem},
 have been a topic of intense investigation in recent years \cite{silbey,recom,thilhs1,thilhs2}. Currently
it is still not clear as to  how  biological systems
comprising hundreds of photosynthetic complexes and many more
correlated excitonic states act in unison to maintain the quantum coherences
in the noisy environment, and attain the much envied high efficiencies at psychological temperatures.

\begin{figure}[htp]
\begin{center}
\subfigure{\label{1a}\includegraphics[width=4cm]{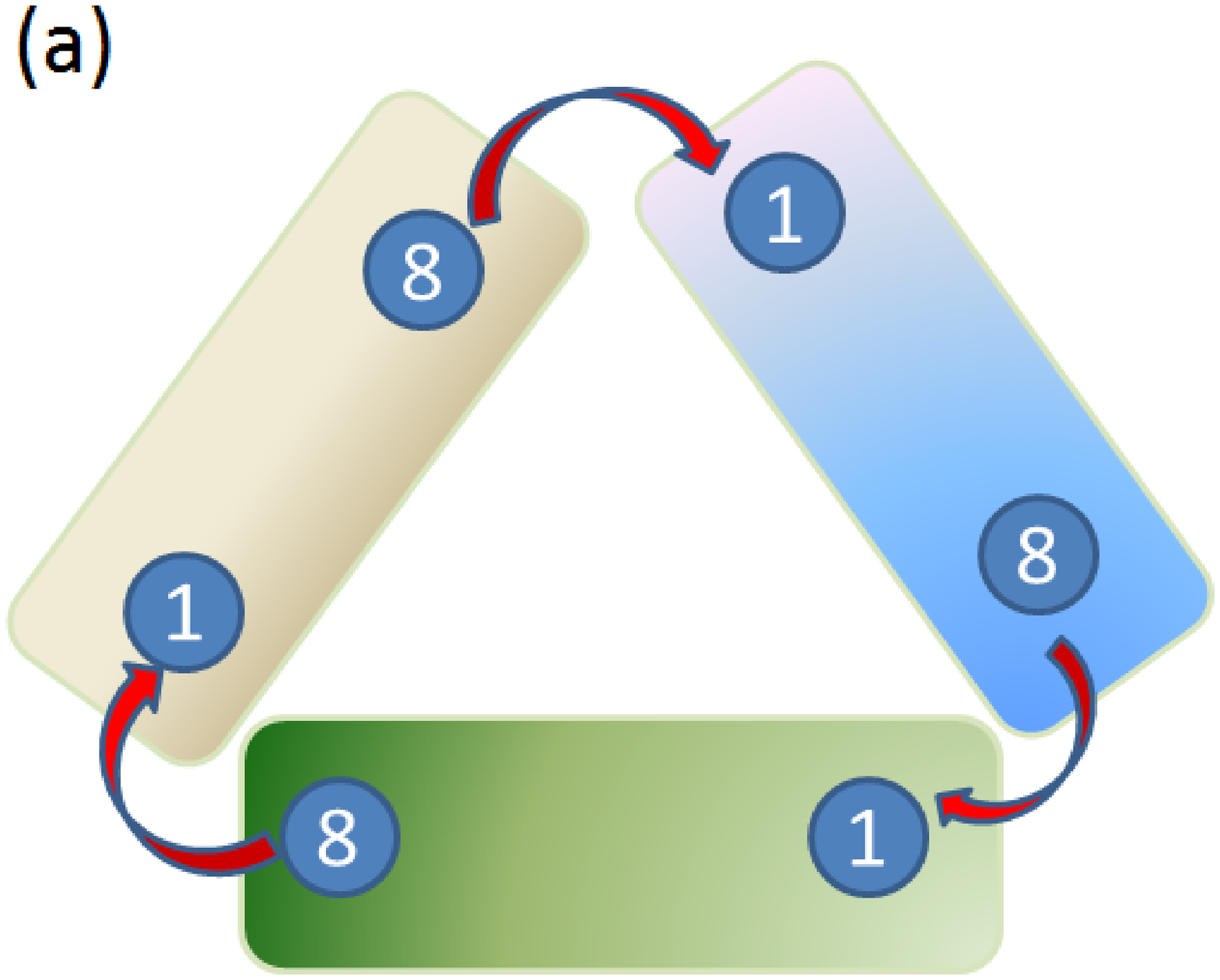}}\vspace{-1.1mm} \hspace{1.1mm}
\subfigure{\label{2b}\includegraphics[width=6.5cm]{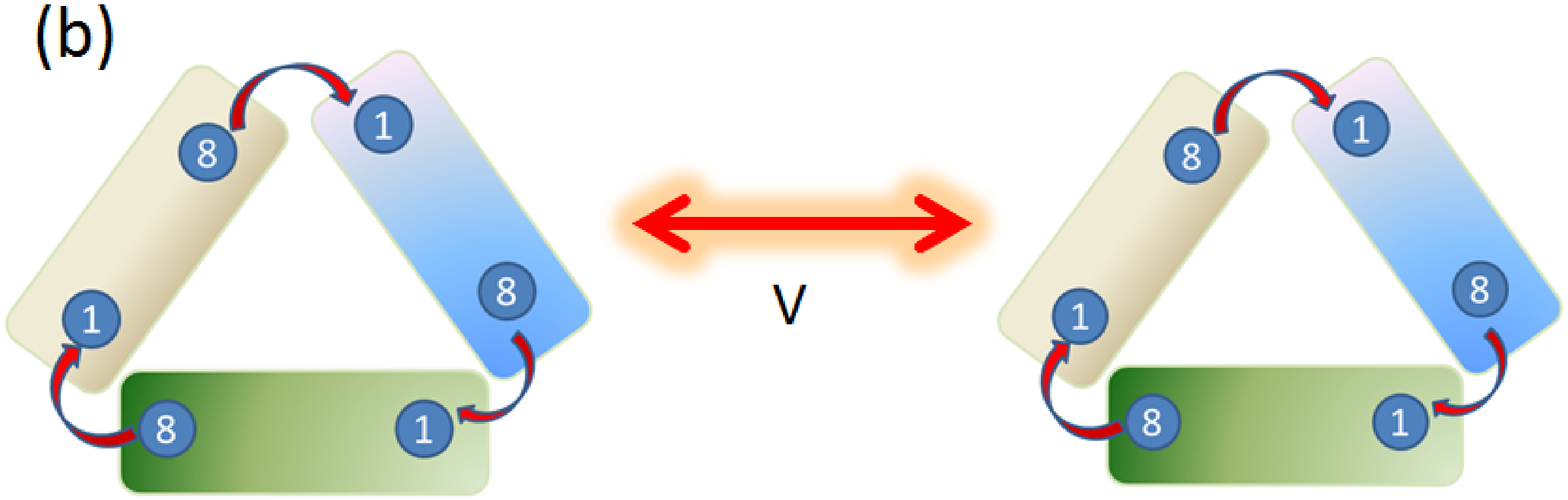}}\vspace{-1.1mm} \hspace{6.1mm}
  \end{center}
\caption{(a) Simplified trimer configuration in which the eighth chromophore is positioned close 
to the first chromophore of a neighbouring monomer, resulting in a 
 stronger inter-monomer (as compared to intra-monomer) interaction for
an  excitation at site $8$ \cite{olb}.
(b)A system of two coupled  FMO trimer  multichromophoric macromolecules that is capable of 
displaying long-lived coherences for small bosonic deviations of composite excitons
at each trimer site.
}
\label{trimer}
\end{figure}

In the FMO complex, the chromophore sites numbered $3$ and $4$ are located  near the reaction center,
and thus are closely linked to the sink region where energy is released,
while chromophore sites $1$ and $2$  are strongly coupled, and dissipate energy via  site $3$ \cite{adolp}.
The sites $1$, $6$, and $8$ are located at the baseplate which connects
to the chlorosomes  that receive electronic excitation. 
Recently, it was shown that  the eighth chromophore (at site $8$) 
is located nearer the chromophore sites of  neighbouring monomers compared to sites in its own monomer \cite{olb}.
This indicates a stronger inter-monomer (as compared to intra-monomer) interaction as far as
the eighth chromophore is concerned and the  excitation at site $8$ is likely 
to propagate to a inter-site monomer (see  Fig.~\ref{trimer}a ). It was pointed out \cite{olb}  that the eigth chromophore 
 acts to facilitate  excitation transfer between  monomers of the FMO trimer even though it is best
placed to receive excitation at the earliest time. To this end, the eighth chromophore
appears to play a critical role in the topological connectivity of
large molecular structures of multichromophoric macromolecule
(MCMM) systems. The electro-optical properties of these MCMMs vary from those
of single chromophores, depending on the  delocalization of excitons within each MCMM.

Here we apply Eq.~(\ref{hermg}) to a 
 model in which excitons in  MCMMs
undergo tunneling dynamics   over  distances that are  large
 compared to the average distance between 
 chromophores within the FMO monomer/trimer configurations \cite{barz}.
A simple  setup is shown in Fig.~\ref{trimer}b, where  $1< N < N_{m}$=24, and which describes the system of
composite bosons in one FMO trimer that is considered as the multichromophoric macromolecule.
Individual MCMM can be coupled to each other as is shown for two FMO trimers in  
Fig.~\ref{trimer}b. Dissipation
may arise from Pauli exchanges between the FMO pigments and the protein bath,
and  recombination and trapping effects specific to each macromolecular system. 
The configuration Fig.~\ref{trimer}b can be further extended to one in which each  MCMM 
includes several trimers forming aggregates. The net dissipation 
within  each MCMM may vary from other similarly configured MCMM,
 depending on the connectivity and proximity to
the region of light illumination. 
In a  recent work \cite{sener},
multiple detrapping/retrapping processes as opposed
to the slow (200 ps)  direct transfer between
RCs in the purple bacterium {\it Rhodobacter
sphaeroides} \cite{sener} were noted  to contribute
to the delocalization of excitation  among several
reaction centres (RCs). In this regard, the tunneling energy $V$ in
Eq.~(\ref{hermg}) may be based
on a cumulative process of repeated trapping/detrapping events
 instead of a single direct transfer mechanism.
 
Oscillations between  MCMMs are expected as  shown in Fig.~\ref{tunnel}a,b,
with excitation exchanges that gradually fades with time depending on the 
 initial conditions, dissipation parameters,
$\gamma_{d1},\gamma_{d2}$ (specific to the two MCMM sites)
and average energy difference ($\approx \omega_0$) between the
  MCMM sites. As noted earlier, coherence times are 
much shorter than the  dissipative  lifetimes  of excitons \cite{flem}.
Of particular interest is  use of the  branching ratio 
in Eq.~(\ref{pas}), which  identifies  effective routes of energy propagation 
in large topologically connected network structures.
A single antenna complex may serve several
FMO complexes, and  a reaction center may be linked to several
FMO trimers. Selective MCMM sites which
function as collection centers may  experience greater dissipation 
than other sites and possess higher branching ratios (Fig.~\ref{branch}a,b). Such sites
contribute to a structural arrangement  that enable photosynthetic organisms to better utilize cellular
resources.

Continued coherence in photosynthetic complexes is  ensured by a small 
bosonic deviation measure $\delta$  and large number $N$  of fermion pairs involved during energy exchanges
(Fig.~\ref{tunnel}). This suggest that a molecular environment  that is highly 
correlated, with a large Schmidt number ${\cal K}$ (Eq.~(\ref{schnum}) is likely
to preserve electronic coherences  needed  for  propagation of  excitation
such that  energy  is  harvested efficiently.
The spectral and molecular dynamics of MCMM  at various
lattice temperatures and excitation densities (light illumination) appear
to influence the kinetics of   exciton and electron-hole pair recombination and relaxation 
processes. These can be seen as critical factors that influence the  Schmidt number ${\cal K}$ specific
to a MCMM. Thus far, we have considered a few factors which underpins
the high efficiencies noted in light-harvesting systems, on qualitative terms.
A quantitative approach would involve modeling
the  realistic condition  of an entire photosynthetic membrane constituting many 
FMO complexes, and taking into account the  topological connectivity
of thousands of bacteriochorophylls. The
approach taken in this work is expected to 
help understand the importance  of the coexistence af boson and fermionic phases,
Pauli scattering effects and selective dissipation in photosynthetic systems.

\section{Conclusion}\label{con}
In conclusion, 
 the quantum dynamics of  conversion of composite bosons 
into fermionic fragment species is demonstrated  using
an open quantum system approach based on 
a system-plus-reservoir model. The  total Hilbert space which
constitutes a  composite boson subspace and an orthogonal fragment subspace  of
 fermions  pairs is  used to examine the effect
of system parameters during the tunneling dynamics 
of coupled composite bosons states. The results highlight 
the interplay  of boson and fermionic phases that is dependent
on density (and indirectly the network connectivity) and lattice temperature,
and   the appearance of exceptional points based on 
experimentally testable conditions (densities, lattice temperatures).
The effect of Pauli exclusion  and other dissipative factors on the 
multichromophoric macromolecules (MCMMs) in photosynthetic light harvesting
is examined in the light of quantitative results obtained for 
coupled composite bosons systems. It is noted that 
long-lived quantum coherence in photosynthetic complexes  is assisted by  small 
bosonic deviation measures (large Schmidt number) and large number  of excitons involved during energy exchanges,
to give rise to a highly correlated  molecular environment. Moreover,
 specific MCMM sites which function as collection centers may possess higher branching ratios,
and  contribute to a structural arrangement  that enable photosynthetic organisms to better 
harness solar  energy  efficiently.

\section{Acknowledgments}

 A. T. gratefully acknowledges the  support of  the Julian Schwinger Foundation Grant,
JSF-12-06-0000, and  thanks M. Combescot for 
useful correspondences on specific properties of composite bosons and related references.

\end{document}